\documentclass[a4paper]{jpconf}
\usepackage{enumerate}
\usepackage{iopams}
\begin{document}
\title{An exact framework for uncertainty quantification in Monte Carlo simulation}
\author{P Saracco$^1$ and M G Pia$^2$}
\address{National Institute for Nuclear Physics (I.N.F.N.) \\ Via Dodecaneso, 33\\ 16146 Genova  (Italy) }
\ead{$^1$ paolo.saracco@ge.infn.it, $^2$ Maria.Grazia.Pia@cern.ch} 
\begin{abstract}
In the context of Monte Carlo (MC) simulation of particle transport Uncertainty Quantification (UQ) addresses the issue of predicting non statistical errors affecting the physical results, i.e. errors deriving mainly from uncertainties in the physics data and/or in the model they embed. In the case of a single uncertainty a simple analytical relation exists among its the Probability Density Function (PDF) and the corresponding PDF for the output of the simulation: this allows a complete statistical analysis of the results of the simulation. We examine the extension of this result to the multi-variate case, when more than one of the physical input parameters are affected by uncertainties: a typical scenario is the prediction of the dependence of the simulation on input cross section tabulations. 
\end{abstract}
\section{Introduction and problem definition}
Uncertainty Quantification (UQ) may refer to a wide variety of specific problems pertaining different scientific areas: related studies involve many methodologies or techniques and even terminologies\cite{g1}-\cite{g10}, so that it is necessary to better delimit the problem: we are interested in studying how uncertainties of the input data needed for physics simulation transfer into the output. Even within this more restricted domain the problem seems quite complex because the meaning of {\em input data needed for physics simulation} is rather undetermined: it may refer to those data needed to specify the experimental setup to be simulated - like, e.g., geometrical data or to the composition of the materials - to some conditions externally applied to the system - like temperature, pressure or electromagnetic fields - or to physical data - cross sections or, often, physical models - needed to describe the transport of particles. All these data are generally affected by uncertainties, which necessarily imply uncertainties in the output of the simulation. So from a responsible MC user's perspective {\em input data} should refer both to those that are under his direct experimental control and to (many) other data, whose values and uncertainties come from different experiments: it is remarkable that often the latter are not considered.  In this contribution we summarize previous results\cite{SarPia2012,SarPia2013} together with some new ones.

To give our analysis a well established mathematically ground we make a fundamental assumption: to be able to disentangle input data uncertainty from the process of simulation itself. In other words we assume that the process of simulation relies on the knowledge of the values of some parameters $x_1,\ldots,x_N$, with their associate uncertainties, and that these values cannot vary during the simulation (or that these values do not depend on the simulation itself): this assumption enables to assume  {\em a priori} the knowledge of the Probability Distribution Function (PDF) $f\left(x_1,\ldots,x_N\right)$ for the input parameters $x_1,\ldots,x_N$, that can then be assumed as stochastic parameters.

This is our main hypothesis: its validity in the generic case must be assessed by users case by case.

We just mention two typical cases under which this hypothesis can be not valid or questionable: the first concerns the local temperature of some experimental apparatus we want to simulate; to some extent it can be assumed as fixed by external conditions with their associate uncertainties, external temperature and/or cooling system of the apparatus if it exists, but local significant temperature variations can be present depending on the local rate of energy deposition due to the transported particles. This last quantity is really tracked by the simulation itself, but its consequence in terms of local temperature variation is not normally taken into account automatically by the MC code: the relation between energy deposition and local temperature variations can be taken into account by coupling the MC simulation with some thermo-mechanical code or model, but even if this can be done, not all MC codes are able to use different cross sections data (cross sections data at different temperatures) in the same simulation. The second case concerns the density of some isotopes, which can vary due to activation/decay processes.

Under the quoted assumption the probability that MC simulation of some physical observable $Y$ have a result between $y$ and $y+dy$ has density
\begin{equation}
g_{MC}(y)\simeq\int_{-\infty}^{+\infty}\,dx_1\cdots dx_N\,f\left(x_1,\ldots,x_N\right)\sqrt{\frac{\displaystyle N_E}{2\pi\sigma_{y_0}^2}}
\exp\left[-\frac{\left(y-y_0\left(x_1,\ldots,x_N\right)\right)^2}{2\sigma_{y_0}^2/N_E}\right]
\label{eqn:PDFSimul}
\end{equation}
for a simulation encompassing $N_E$ events. Here $y_0\left(x_1,\ldots,x_N\right)$ is the sampled mean for the physical observable $Y$ when the input unknowns assume values $\left(x_1,\ldots,x_N\right)$ and $\sigma_{y_0}^2$ is its sampled variance. This result derives from the Central Limit Theorem (CLT) if $N_E$ is sufficiently large to make $\sigma_{y_0}^2$ independent from $N_E$ itself. In the limit $N_E\to\infty$
\begin{equation}
g(y)=\int_{-\infty}^{+\infty}\,dx_1\cdots dx_N\,f\left(x_1,\ldots,x_N\right)\delta\left(y-y_0\left(x_1,\ldots,x_N\right)\right)\,.
\label{eqn:PDFExact}
\end{equation}
Equation (\ref{eqn:PDFExact}) mathematically states how uncertainties in the input data {\em exactly transfer} into our knowledge of the observable $Y$ because this is an assignment of a probability for each possible outcome of this observable.

Even if we derived (\ref{eqn:PDFExact}) from (\ref{eqn:PDFSimul}) we could as well make the opposite: (\ref{eqn:PDFExact}) represents the forward propagation of uncertainty for a deterministic problem having solution $y_0$ which, in turn, depends (parametrically, not dynamically) on some unknown input values whose PDF is given by $f\left(x_1,\ldots,x_N\right)$. Searching the solution of such deterministic problem by MC simulation - that is we go back from (\ref{eqn:PDFExact}) to (\ref{eqn:PDFSimul}) - results in a (gaussian) statistical blurring of the exact result $y=y_0\left(x_1,\ldots,x_N\right)$ of the underlying deterministic problem. This dual interpretation of (\ref{eqn:PDFExact}) and (\ref{eqn:PDFSimul}) is crucial to our investigation: it is a clear indication of which is the natural goal for any UQ project - the determination of $g(y)$ - and of its prerequisite - the knowledge of $f\left(x_1,\ldots,x_N\right)$. Unfortunately, but quite obviously, the determination of $g(y)$ requires also the knowledge of the parametric dependence of the exact solution  $y=y_0\left(x_1,\ldots,x_N\right)$ on the input uncertainties and on their PDFs: except for very simple examples we are not able to exploit this dependance, but clearly we can use simulation to extract these required informations, as we shall see. It is relevant to stress that our attention has turned to the determination of the exact form of $g(y)$ rather than the one of $g_{MC}(y)$: with this passage we obtain an important result both on the conceptual - as we have seen - and on the practical side. In fact the inherent dependency of $g_{MC}(y)$ from the details of the simulation\footnote{From (\ref{eqn:PDFSimul}) the dependence on the number of generated events $N_E$ is explicit, but obviously $g_{MC}(y)$ depends also on the specific way MC simulation is implemented. } makes it impossible to determine it other than by statistical sampling: this implies the necessity of running a very large number of simulations, each time sampling $(x_1,\ldots,x_N)$ out of $f(x_1,\ldots,x_N)$. On the contrary the determination of the dependence  $y_0=y_0\left(x_1,\ldots,x_N\right)$ can be accurately obtained with fewer runs, as we shall discuss. 

Uncertainty quantification directly derives from (\ref{eqn:PDFExact}): in fact if we know $g(y)$ the determination of any required statistical information of the physical output, e.g. its confidence intervals, from the unknown input parameters is straightforward. So the task of any UQ project is not the study of how uncertainties of the input data needed for physics simulation transfer into the output (of the simulation), that is the determination of $g_{MC}(y)$, rather the use of simulation to determine the properties of $g(y)$ as better as possible. We note, {\em en passant}, that techniques we are going to develop to study $g(y)$ are then applicable to more general contexts than simulation, for instance also when simulations are coupled to deterministic codes.

Moreover we learn that the necessary prerequisite is a precise, as far as possible, knowledge of the PDF for the input data, because it is directly transferred into the physical output probability distribution. 

In most of the cases uncertainties in the data which are under the user's control are independent from uncertainties in the physical data needed for transport. So our approach enables to clearly distinguish two very different phases of UQ projects: (i) a {\em validation phase} and (ii) a {\em problem specific analysis phase}.
\begin{enumerate}[(i)]
\item {\bf The validation phase} -
All physical data needed for transport should be validated: all general purpose MC codes make use of cross section tabulations as well as of physical models. All these data should be carefully analyzed giving to users at least the knowledge about their confidence intervals (better, their PDF). It should be clear that this cannot be a users's responsibility: without a previous validation phase any UQ project is meaningless. On the other side how much these data are individually relevant in any given problem cannot be a priori established. 
\item {\bf The problem specific analysis phase} - Most of realistic situations under simulation may involve {\em a priori} hundreds of physical parameters in their full definition: they include parameters whose confidence intervals (or PDF) come from the validation phase, as well as problem specific parameters, like, e.g., geometrical or material composition parameters: for these last is user's responsibility to establish at least proper confidence intervals, better PDF. It will become clear that a full UQ process is out of human possibilities and largely meaningless. So a very detailed analysis of the problem at hand should be carried on to identify those parameters which are likely to be the most relevant in the given experimental configuration. Methods we shall expose in Section \ref{sect:ThePath} can be applied to the selected set of parameters. If necessary successive iterations must be carried on other parameters.
\end{enumerate}

\section{The path to UQ\label{sect:ThePath}}
Once we have selected a set of parameters on which perform a UQ (or sensitivity) analysis we can proceed on the basis of 
(\ref{eqn:PDFExact}) and (\ref{eqn:PDFSimul}): in most of the cases two further assumptions hold, namely (i) variations of the input parameters are independent and (ii) $y_0$ is linear in each of the parameters in their range of variability. We will discuss later these assumptions. If this is true (\ref{eqn:PDFExact}) simplifies greatly to
\begin{equation}
g(y)=\int_{-\infty}^{+\infty}\,dx_1\cdots dx_N f_1(x_1)\cdots f_N(x_N)\delta\left[y-\bar y_0-\sum_{k=1}^N
a_k(x-\bar x_k)\right]
\label{eqn:PDFLinear}
\end{equation}
where $f_j(x_j)$ and $\bar x_j$ can be assumed as known from validation phase or from problem specific analysis phase. Problem defined by (\ref{eqn:PDFLinear}) is a completely defined mathematical problem once we are able to determine $\bar y_0$ and $a_j=\left.\frac{\displaystyle\partial y_0\left(x_1,\ldots,x_N\right)}{\displaystyle\partial x_j}\right\vert_{x_k=\bar x_k}$: it is obvious that we can obtain a statistical estimate of these required values with a predetermined accuracy by means of $N+1$ simulations each run with a different set of input parameters, namely $\bar x_1,\ldots,\bar x_N$ and  $\bar x_1,\ldots,\bar x_j+\Delta,\ldots,\bar x_N$, so that indicating with $y_{MC}\left(x_1,\ldots,x_N\right)$ the output of the simulation with a given set of values for the input parameters
\begin{eqnarray}
\bar y_0 &=& y_{MC}\left(\bar x_1,\ldots,\bar x_N\right)\nonumber\\
a_j &=& \left[y_{MC}\left(\bar x_1,\ldots,\bar x_j+\Delta,\ldots,\bar x_N\right)-y_{MC}\left(\bar x_1,\ldots,\bar x_N\right)
\right]/\Delta\nonumber\,.
\end{eqnarray}
Obviously these values are affected by statistical errors of the simulation that are of the order of $\sigma_{y_0}/\sqrt{N_E}$\footnote{We can naturally assume $\sigma_{y_0}$ as constant for small variations $\Delta$ of the parameters.}: then a rough estimate of the number of events needed in each of the simulations comes from
\begin{equation}
\nonumber \sigma_{y_0}/\sqrt{N_E}\ll \left\vert y_{MC}\left(\bar x_1,\ldots,\bar x_j+\Delta,\ldots,\bar x_N\right)-y_{MC}\left(\bar x_1,\ldots,\bar x_N\right)\right\vert
\label{eqn:NECond}
\end{equation}
It should be realized that the use of (\ref{eqn:PDFLinear}) entails a large reduction in the computer time required with respect to any attempt to directly determine $g_{MC}(y)$, a task that would require thousands of MC runs.

From (\ref{eqn:PDFLinear}) it is easy to extract $<y>$ and $<y^2>$; if for instance $\bar x_k=<x_k>$, often a convenient choice
\begin{eqnarray}
<y>&=&\bar y_0\nonumber\\ \nonumber
<y^2>&=&\bar y_0^2+\sum_{k=1}^N a_k^2<(x_k-\bar x_k)^2>=\bar y_0^2+\sum_{k=1}^N a_k^2\sigma_{k}^2
\end{eqnarray}
so that $\sigma^2_y=\sum_{k=1}^N a_k^2\sigma_{k}^2$. This quantity can be assumed as a first estimate of the uncertainty, but this is really correct only if $g(y)$ is - at least approximately - a normal distribution: this is certainly true if  the $f_j$ are normal distributions as well, but not in other cases. So we are left with a classical problem in probability theory, namely the determination of the distribution for the weighted sum of $N$ independent stochastic variables obeying (a priori) arbitrary distributions.

In some cases the calculation of $g(y)$ can be carried on exactly. We quoted the case when all the $f_j$s are normal distributions: this is a specific example of a general class of distributions which {\em characteristic functions} - the Fourier transform of the given distributions - are closed under product: these are the so-called {\bf stable distributions}. In probability theory, a random variable is said to be stable (or to have a stable distribution) if it has the property that a linear combination of two independent copies of the variable has the same distribution, up to location and scale parameters. The stable distribution family is also sometimes referred to as the {\em L\'evy} \cite{Levy1925} {\em $\alpha$-stable distribution}. Distributions belonging to this family \cite{Zolo1980,FoNo1999} have characteristic functions of the form
\begin{equation}
\phi(q;\mu,c,\alpha,\beta)=\exp\left[i t\mu-\left\vert c t\right\vert^\alpha\left(1-i\beta{\rm sgn}(t)\right)\Phi\right]
\end{equation}
where $\Phi=\tan(\pi\alpha/2)$ if $\alpha\not=1$ and $\Phi=-2\log\vert t\vert/2$ if $\alpha=1$. This is a 4 parameters ($-\infty<\mu<\infty, 0<c<\infty,0<\alpha\le 2$ and $-1\le\beta\le 1$) family which is closed under product for fixed $\alpha$: the analytic form of the corresponding PDF is known only for some special values of the parameters. Among these we mention the normal distribution ($\alpha=2$, $c=\sigma/\sqrt 2$), the Cauchy distribution ($\alpha=1$, $\beta=0$), and the L\'evy distribution ($\alpha=1/2$, $\beta=1$). All these distributions apart the normal one are {\em heavy tailed}, that is their behavior for large 
$x$ is $\vert x\vert^{-1-\alpha}$. In the limit $\alpha\to 0$ or $c\to 0$ they approach a Dirac delta.

The most common case of applicability of these properties to UQ is when we have to analyze a certain number of input parameters affected by different {\em statistical errors}. 

Another useful case in practice is when we have a certain number of input parameters with flat distributions: this is the case when input parameters are given in the form $\bar x_j\pm\Delta x_j/2$. In this case $g(x)$ is given by a generalization of the {\em Irwin-Hall distribution} \cite{Irw1927,Hall1927} recently revisited \cite{BraGu2020}.

Unfortunately all these results apply only to cases when all the input unknowns have the same distributions with different parameters: for instance they can be all normal distributions, with different means and variances, or all flat distributions, with different means and experimental errors, and so on. It should be clear that in a generic UQ problem this is a strong limitation, because we can have the case of input parameters obeying different distributions as well: this problem is clearly not soluble in full generality so we must work on some approximation scheme to the search for $g(y)$.

\section{A more general result}
We recently succeeded in finding an exact analytical expression for the weighted sum of $N$ independent stochastic variables obeying arbitrary polynomial distributions supported on bounded intervals, whose proof can be found in \cite{SarPia2013a}. Here we simply quote our central result and we show how this can be useful in determining $g(y)$ with some predetermined error: an arbitrary distribution $f(x)$ supported on a bounded interval $a\le x\le b$ can always be approximated with a predetermined accuracy by a sequence of polynomials each defined on different sub-intervals of  $[a, b]$: the most simple case of such procedure is to subdivide the interval $[a, b]$ in $k$ sub-intervals $x_0=a,x_1,\ldots x_k=b$ and to approximate the given distribution with sequence of segments joining the values $f\left(x_0\right),\ldots f\left(x_k\right)$. More accurate results can be obtained with a lower number of subintervals by using splines. Accuracy of an approximation is defined with respect to some given norm, measuring the {\em distance} between the exact result and its approximation; in our case, to the purpose of studying the propagation of error, it is convenient to make the choice
\begin{equation}
\nonumber ||f||={{\rm sup}\atop{[a, b]}}\vert f(x)\vert
\end{equation}
so that the distance between two function is the maximum of their absolute differences. We assert without proof that if $f$ is continuous (or if it has at most a finite number of discontinuities) it is always possible to find a subdivision of $[a, b]$ such that the distance between $f$ and its polynomial approximation $f_{\rm app}$ is bounded by a predefined $\varepsilon>0$\footnote{This is a consequence of Weierstrass approximation theorem.}. Then if we have $N$ such distributions, each approximated by this procedure, the maximum error of the distribution of the sum $x_1+\ldots+x_N$ is clearly $N\varepsilon$: for a weighted sum $a_1 x_1+\ldots+a_N x_N$ the bound on the error turns out to be $\varepsilon\displaystyle \sum_j \left\vert a_j\right\vert$.   This means that with this procedure we can obtain the required distribution $g(y)$ with any predetermined error provided we are able to make exactly the convolution\footnote{We remind the reader that the (weighted) sum of $N$ stochastic variables is expressed by the convolution of their (rescaled) distributions.} of generic polynomial forms defined over different intervals. 

Thanks to the linearity of the convolution this is equivalent to the ability of performing the convolution of $N$ monomial forms with arbitrary exponents supported on different bounded intervals: this is exactly what we proved in \cite{SarPia2013a}. The convolution of $N$ monomials $x_1^{p_1},\ldots,x_N^{p_N}$ that are different from zero only over the intervals $-c_1\le x_1\le c_1$, $\ldots$, $-c_N\le x_N\le c_N$ is given by
\begin{eqnarray}\label{eqn:GenConv}
I(x;\vec c;\vec p;N)&=&\frac{\displaystyle \prod_{k=1}^N({\rm max}(p_k,1))!}{\displaystyle 2\left(\sum_{k=1}^N(p_k+1)-1\right)!}
\prod_{k=1}^N\hat Q\left(p_k,\lambda_k\right)\\ &&\times\left.
\sum_{m\in{\mathbb M}_N(x,\vec c,\vec\lambda)}(-1)^{C(m)}m^{\sum_{k=1}(p_k+1)-1}
{\rm sgn}\left(m\vert_{\vec\lambda=1}\right)\right\vert_{\vec\lambda=1}
\nonumber
\end{eqnarray}
where $\displaystyle \hat O(q,\lambda)=\sum_{s=0}^q\frac{\displaystyle (-1)^s}{\displaystyle s!}
\frac{\displaystyle\partial^s}{\displaystyle\partial\lambda^s}$ and
${\mathbb M}_N(x,\vec c,\vec\lambda)=\left\{m(x)=x\pm c_1\lambda_1\pm c_2\lambda_2\ldots\pm c_N\lambda_N\right\}$.
Equation (\ref{eqn:GenConv}) appears as a generalization of the result \cite{BraGu2020}. We remark that calculation of $g(y)$ along the quoted guidelines can be tedious and often long, so we are planning to release a specific dedicated software to automate the process. 

Anyway the outlined procedure {\bf is always able to give a definite answer on the expected PDF} $g(y)$ with a predefined accuracy: it is then possible to extract confidence intervals for the required physical quantity, or any other statistical information we can need.

Two sources of errors must be accurately followed up: the first concerns the (statistical) errors in the values required for the calculation ($\bar y_0$ and the $a_j=\displaystyle\left\vert\frac{\partial y_0}{\partial x_j}\right\vert_{x_k=\bar x_k}$) that, as we told, can be extracted from accurate simulations; the second concerns propagation of errors in the procedure of approximating individual input PDFs with polynomial forms over finite intervals. 

In the most generic case some of the input PDF can be defined over infinite intervals - for instance they may be normal distributions: if this is the case we can follow the described procedure provided we {\em a priori} define some confidence for the whole calculation; that is to say that a normal distribution can be approximated by a distribution over a finite interval with a given confidence. So the worst case, for which we cannot easily apply the described procedure, is the case when some of the input PDF are heavy tailed.

\section{Conclusions and outlook}
Results presented in the previous section provide a well defined mathematical framework to obtain the probability density function for any observable $Y$ depending on some given set of input physical parameters and on their associate uncertainties. 

We made three assumptions about which some comment is useful.

First we assumed that the values of the input parameters, and implicitly their probability distributions, do not vary along the simulation: on the basis of the discussion we made about equations (\ref{eqn:PDFSimul}) and (\ref{eqn:PDFExact}) this condition should be more properly reformulated as: the input parameters and their probability distributions do not depend on the dynamical evolution of the system. It should be clear that this condition can, at least in principle, be given up: this possibility depends on our ability to build a more complete dynamical description of the system under examination including also the possible evolution of these dynamically variable parameters: to remain within the examples we made, local temperatures can be derived by some thermomechanical or thermohydraulic model which must be coupled to particle transport, or isotope concentrations can be obtained by coupling transport to some solver for Bateman's equations; even if this is in principle possible, in practice it depends on a major reformulation of the existing simulation codes.

The second major assumption is about the statistical independence of different input uncertainties: we currently do not see any way-out to this assumption; however this is the most common case in practice.

The third assumption is about the linearity of the response $y_0(x_1,\ldots,x_N)$ with respect to each of the $x_k$, see equation (\ref{eqn:PDFLinear}): it is clearly possible to give up this assumption by properly subdividing the region of integration in (\ref{eqn:PDFExact}), as we noted earlier, for example using methods from \cite{LuSte2004}, but this process introduce additional approximation errors whose propagation must be studied case by case for their effect on the precision on the knowledge of $g(y)$.

So we can conclude that the presented conceptual mathematical framework for UQ is really consistent and it relies in principle only on the assumption of mutual independence of the input unknowns: however its practical usability is not straightforward in its general formulation because of the error tracking process. Then we plan the release of a dedicated software tool to automate this task.

\end{document}